\documentclass[twocolumn,prl,aps,nobalancelastpage]{revtex4}

\usepackage{graphicx}

\begin{document}

\title{Determination of the Fermi Surface of MgB$_2$ by the de Haas-van Alphen effect}

\author{A. Carrington$^a$, P.J. Meeson$^a$, J.R. Cooper$^b$, L. Balicas$^c$,  N.E. Hussey$^a$,  E.A. Yelland$^b$,
 S. Lee$^d$, A. Yamamoto$^d$, S. Tajima$^d$,  S.M. Kazakov$^e$ and J. Karpinski$^e$}

\affiliation{$^a$H. H. Wills Physics Laboratory, University of Bristol, Tyndall Avenue, BS8 1TL, United Kingdom.}

\affiliation{$^b$Interdisciplinary Research Centre in Superconductivity and Department of Physics, University of
Cambridge, Madingley Road, Cambridge CB3 0HE, United Kingdom.}

\affiliation{$^c$National High Magnetic Field Laboratory, Florida State University, Tallahassee, Florida 32306, U.S.A.}

\affiliation{$^d$Superconductivity Research Laboratory, International Superconducting Technology Center, Tokyo 135-0062, Japan.}%

\affiliation{$^e$Laboratorium f\"{u}r Festk\"{o}rperphysik, ETH Z\"{u}rich, CH-8093 Z\"{u}rich, Switzerland.}

\date{\today}
\begin{abstract}
We report measurements of the de Haas-van Alphen effect for single crystals of MgB$_2$, in magnetic fields  up to 32
Tesla. In contrast to our earlier work, dHvA orbits from  all four sheets of the Fermi surface were detected. Our
results are in good overall agreement with calculations of the electronic structure and the electron-phonon mass
enhancements of the various orbits, but there are some small quantitative discrepancies. In particular, systematic
differences in the relative volumes of the Fermi surface sheets and the magnitudes of the electron-phonon coupling
constants could be large enough to affect detailed calculations of T$_{c}$ and other superconducting properties.
\end{abstract}

\maketitle

There has been rapid progress in  understanding the physical properties of magnesium diboride since the discovery of
superconductivity at 39 K just over two years ago. The accepted consensus is that it is an $s$-wave, phonon-mediated
superconductor but with some highly unusual properties.  The most important of these are the anomalously high $T_c$ and
the existence of two, almost distinct, superconducting gaps. This understanding is based on many experiments and on
theoretical calculations  of  the unusual electronic structure of MgB$_2$ \cite{kortus01,liu01,choi02,mazin02}.

\begin{figure}
\includegraphics[width=7.0cm,keepaspectratio=true]{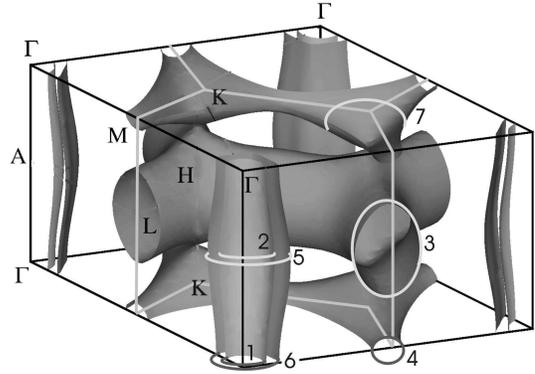}
 \caption{Calculated Fermi surface topology, with possible dHvA extremal orbits (for frequencies $<$10~kT) indicated.
  Figure adapted from Kortus \textit{et al.} \protect\cite{kortus01}. }
 \label{figbs}
\end{figure}

There have been two  direct experimental probes  of the Fermi surface (FS)  structure of MgB$_2$; angle-resolved
photoemission spectroscopy \cite{uchiyama} and the de Haas-van Alphen effect (dHvA)\cite{yelland02}. Our previous dHvA
study \cite{yelland02} showed good agreement between theory and experiment regarding the areas of the orbits that were
observed and their electron-phonon mass enhancements. However, only three out of nine predicted dHvA orbits were
observed (labelled 1,2 and 3 in Fig.~\ref{figbs}) and so information was only obtained about two of the four FS sheets
predicted by band calculations. It was not entirely clear whether the unobserved orbits were missing simply because of
their relatively short mean-free-paths, or whether the topology of the other sheets was substantially different. This
issue clearly affects calculations of many physical properties (especially $T_c$). In this Letter, we report new
measurements of the dHvA effect in magnetic fields up to 32~Tesla for two single crystals. These new results give
evidence for orbits on all four sheets of the Fermi surface.

Experiments were conducted on two different crystals grown by different groups.  Sample B is one of the crystals
studied previously \cite{yelland02} and was grown in Tokyo by  high pressure synthesis  using natural boron. Sample K
was grown in Z\"{u}rich by a similar technique but isotopically pure boron-10 was used as a starting ingredient
\cite{b10note}. Other crystals from the same batches were found to have $T_c$ (onset) of 38.0~K and 37.7~K for the
Tokyo and Z\"{u}rich crystals respectively. dHvA oscillations were observed by measuring the torque ($\Gamma$) with a
sensitive piezo-resistive cantilever technique \cite{bergemannphd}. Both samples were studied extensively at Bristol in
fields up to 20.5~T and temperatures down to 0.3~K, and in fields up to  32~T  at the NHMFL.

The first harmonic of the oscillatory part of the torque for a 3D Fermi liquid is given by \cite{shoenberg,wasserman}
\begin{equation}
\Gamma_{osc}\propto\frac{1}{C} \frac{dF}{d\theta}B^{\frac{1}{2}}
R_D R_T R_S \sin \left(\frac{2\pi F}{B}+\gamma\right)
\label{lkexpression}
\end{equation}
where $F$ is the dHvA frequency [$F = (\hbar/2\pi e)A$,  $A$ is the extremal orbit area in $\mathbf{k}$-space]; $C$ is
the curvature factor, $\gamma$ is the phase; $R_D$, $R_T$ and $R_S$ are the damping factors from impurity scattering,
temperature and spin splitting respectively.  The Dingle factor, $R_D = \exp(-\frac{\pi m_B }{eB\tau})$, where $m_B$ is
the unenhanced band mass \cite{shoenberg,wasserman} and $\tau$ is the scattering time. An equivalent expression for
$R_D$ is $R_D=\exp(-\frac{\pi \hbar k_F}{ eB\ell})$, which shows clearly the increased damping with shorter
mean-free-path $\ell$ and increased average Fermi wavevector ($k_F$) of the orbit. $R_T = X/(\sinh X)$ where $X =
\frac{2\pi^2k_{_B}}{\hbar e} \frac{m^*T}{B}$, $m^*$ is the quasi-particle effective mass. Finally, the spin splitting
factor is given by $R_S = \cos(\frac{ \pi n g m_B(1+S)}{2m_e})$ where $(1+S)$ is the orbitally averaged
exchange-correlation (Stoner) enhancement factor, $g$ is the electron $g$-factor, $m_{e}$ is the free-electron mass and
$n$ is an integer.

The electronic structure and dHvA orbits of MgB$_2$ have been calculated by three different groups
\cite{harima,mazin02prb,rosner02}.  The calculated Fermi surface is shown in Fig.~\ref{figbs}, together with the
expected dHvA extremal orbits \cite{harimanote}.  The calculations \cite{kortus01} show that the electronic states near
the Fermi level arise primarily from the boron atomic orbitals. The calculated Fermi surface is composed of  four
distinct sheets. Two of these arise from boron  $\sigma$ orbitals and are  quasi-two dimensional warped cylinders
running along the $c$-direction, whereas the other two are tubular networks with larger  $c$-axis dispersion, that are
mainly formed from the boron $\pi$ orbitals. In total 9 primary extremal orbits have been predicted, and 7 of these are
labelled in the Figure (two orbits with frequencies $>$30kT have been omitted). Calculations of the dHvA frequencies
and masses of  the various  orbits by the three groups are all in good agreement (the differences are typically
100-200~T, and $\lesssim 5$\% respectively).

\begin{figure}
\includegraphics[width=8.0cm,keepaspectratio=true]{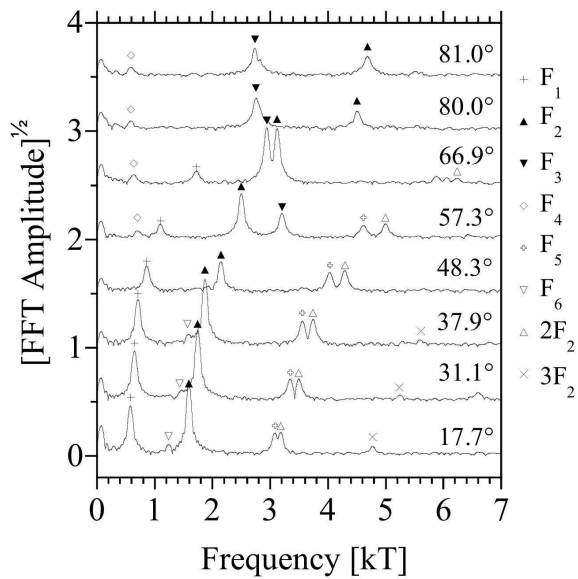}
 \caption{Fourier transforms of the raw high field data for crystal K at 1.4~K and fields from 20$\rightarrow 32$~T.
 (the square root of the FFT amplitude is plotted and the data are offset for clarity).
 The orbit assignments are indicated by symbols and follow the notation of Figure \protect\ref{figbs}}
 \label{fftfig}
\end{figure}

The fast Fourier transforms (FFTs) of the raw torque data between 20 and 32 T at 1.4 K for crystal K are shown in Fig.
\ref{fftfig}, as a function of angle $\theta$ as the sample was rotated from $H\|c$ to $H\|a$ ($H\|c \equiv 0^\circ$).
In addition to the frequencies observed in our previous study \cite{yelland02} ($F_1$, $F_2$ and $F_3$) several
additional peaks are visible. Some care is needed in interpreting these new frequencies. Because the cantilever is
deflected slightly by the torque, the torque measurements are not made at constant angle. This generates spurious
harmonics  and combinations of the main dHvA oscillations. For crystal K only weak harmonics of F$_2$ are
 observed (just above $F_5$-- see Fig.\ \ref{fftfig}), whilst for the larger crystal B, several harmonics of $F_3$
and a frequency corresponding to $F_1 + F_2$ were observed.

\begin{figure}
\includegraphics[width=7.0cm,keepaspectratio=true]{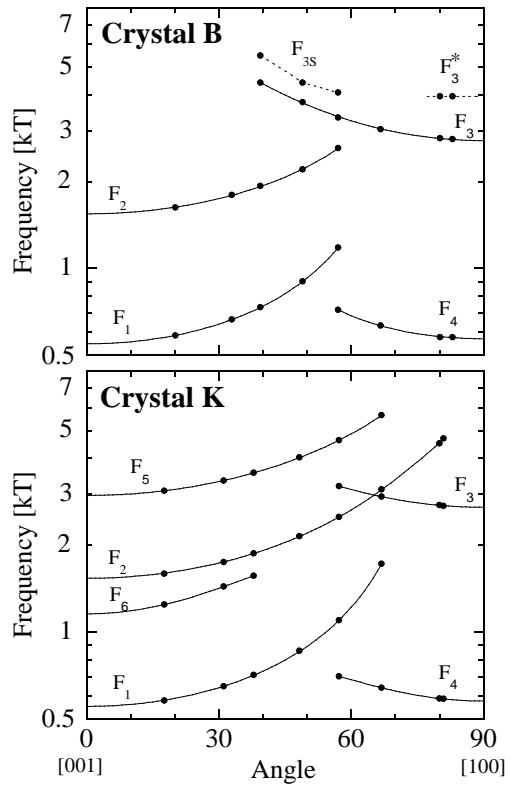}
 \caption{Observed frequencies versus angle as the samples were rotated from $H \|c$
 to (approximately) $H \|a$.  The solid lines are polynomial fits to the data as described in the text,
 and the dotted lines are guides to the eye.}
 \label{figrot}
\end{figure}

\begin{table*}
\caption{Summary of dHvA parameters for both samples (K and B) along with the theoretical predictions
(Th)\protect\cite{mazin02prb}}
\begin{center}
\begin{tabular}{cc|ccccccccc|cccc}
\hline\hline
&&\multicolumn{9}{c}{\bf Crystal K}&\multicolumn{4}{|c}{\bf Crystal B}\\
 Orbit&&$~F^0_{\rm exp~}$&$~F^{~}_{\rm Th}~$&$\Delta F$&$~\Delta E$&~$\ell$~&~$m^*_{\rm exp}$~&~$m_{_{\rm
band}}$~&$\lambda^{\rm ep}_{\rm exp}$&~$\lambda^{\rm ep}_{\rm Th}$~&~$F^0_{\rm exp~}$&~$\ell$~&~$m^*_{\rm exp}$~&$\lambda^{\rm ep}_{\rm exp}$\\
& &[T]  & [T]& [T] & [meV] & [\AA] & [$m_e$] &[$m_e$]&~&~&[T]&[\AA]&[$m_e$]&~\\
\hline
1&$\sigma$&551&730&$+179$&$+83$&550&$0.548\pm0.02$&0.251&$1.18\pm0.1$&1.25&546&380&$0.553\pm0.01$&$1.20\pm0.04$\\

2&$\sigma$&1534&1756&$+222$&$+82$&900&$0.610\pm0.01$&0.312&$0.96\pm0.03$&1.25&1533&580&$0.648\pm0.01$&$1.08\pm0.03$\\

3&$\pi$&2705&2889&$+184$&$-67$&570&0.439$\pm0.01$&0.315&$0.40\pm0.03$&0.47&2685&680&$0.441\pm0.01$&$0.40\pm0.03$\\

4&$\pi$&576&458&$-118$&$-56$&$-$&$0.31\pm0.05$&0.246&$0.31\pm0.1$&0.43 &553&$-$&$0.35\pm0.02$&$0.42\pm0.08$\\

5&$\sigma$&2971&3393&$+422$&$+79$&390&$1.18\pm0.04$&0.618&$0.91\pm0.07$&1.16&$-$&$-$&$-$&$-$ \\

6&$\sigma$&1180&1589&$+409$&$+87$&$-$&$1.2\pm0.1$&0.543&$1.2\pm0.2$&1.16&$-$&$-$&$-$&$-$ \\

\hline\hline
\end{tabular}
\end{center}
\label{tablesum}
\end{table*}

The observed frequencies (omitting those assigned as harmonics or combinations) are shown in Fig.\ \ref{figrot} as a
function of $\theta$ for both crystals. The solid lines in Fig.\ \ref{figrot} are fits of the observed $F(\theta)$
values to $F(\theta) = \sum_{i=1}^3 a_i/\cos^i(\theta-\theta_0)$ ($\theta_0=0$ or $90^\circ$), which we use to
extrapolate the observed frequencies to the symmetry points (we denote these frequencies as $F_n^0$). The assignment of
the FFT peaks to the orbits shown in Fig.\ \ref{figbs} was achieved by comparing the values of the frequencies obtained
by extrapolation, and their angular dependencies, with the calculations.

For crystal K, signals from 6 orbits associated with all 4 sheets of Fermi surface are observed.  We are therefore able
to verify experimentally the  Fermi surface topology predicted by Kortus \textit{et al.} \cite{kortus01} and shown in
Fig.\ \ref{figbs}. For crystal B, in addition to the 3 orbits observed previously, 3 further frequencies are seen. One
of these can be assigned to $F_4$. The frequency of $F_3^*$ (F=4600~T) is close to that predicted for orbit 7
(F=4294~T)\cite{harima} (and has a similar effective mass), however, the frequency is also close to that expected for
an orbit equivalent to $F_3$ but from the tube oriented along $a^*$ (i.e., at $60^\circ$ to $a$). A subsequent in-plane
rotation study showed the latter to be the most likely origin. The origin of $F_{3S}$ is less clear, but it could arise
from a slight warping of the same in-plane tube responsible for $F_3$.

Table \ref{tablesum} shows that the values of $F^0$ found for the two crystals are in good agreement. In total, we have
studied crystals from six batches (5 from Tokyo and one from Z\"{u}rich)  and so far, the dHvA frequencies agree to
within 30~T or 0.06 \% of the basal area of the first Brillouin zone (50.2~kT). This suggests that although $T_c$ is
slightly reduced compared with the best polycrystalline samples, any possible Mg deficiency is very reproducible, and
probably small \cite{cooper03}.

The differences between the measured $F^0$ values and those predicted by theory  is a significant fraction of $F^0$ in
some cases, but we note that 100~T discrepancy only amounts to $\sim$ 0.3 \% of the basal area of the Brillouin zone.
The volumes of the tubes are proportional to the average of the two extremal areas \cite{rosner02} and these are both
$\sim 16$\% smaller than the calculations \cite{mazin02prb}, implying a corresponding reduction in the number of holes
in these two tubes \cite{rosner02}. This may have a significant effect on calculations of physical properties such as
$T_{c}$, its dependence on alloying or pressure, and the London penetration depth.

It is instructive \cite{mazin02prb,rosner02} to calculate the Fermi energy shift $\Delta E$ which would bring the
theoretical frequencies in line with experiment.  As the dHvA band mass is defined as
$m_B=\frac{\hbar^2}{2\pi}\frac{\partial A}{\partial E}$, the necessary band shifts are given by $\Delta E = \frac{\hbar
e }{m_B} \Delta F$ (where $\Delta F = F_{\rm Th}-F_{\rm exp}$). There is remarkable consistency between the $\Delta E$
values for the orbits, with the values roughly falling into two groups (Table \ref{tablesum}). For the $\sigma$ sheet
orbits (1,2,5,6) the average shift is $83\pm 4$ meV, whereas for the $\pi$ sheet orbits (3,4) it is $-61\pm 5$ meV.
Because of the high degree of reproducibility of the frequencies between samples, it is unlikely that this discrepancy
is caused by sample impurities or non-stoichiometry.  Instead, it seems to imply a shortcoming of the LDA
calculations\cite{mazin02prb,rosner02}.

Quasiparticle effective masses were determined by performing field sweeps at different temperatures and fitting the
amplitudes to Eq.\ \ref{lkexpression}.   Our experimental values of $m^*$ are compared with the calculated band masses
in Table\ \ref{tablesum}.  In MgB$_2$ the dominant source of mass enhancement is the electron phonon interaction. If we
assume this is the only source of enhancement we can calculate an upper bound for the electron-phonon coupling
constants $\lambda^{\rm ep}$, from $\lambda^{\rm ep}= m^*/m_B -1$. The results (Table \ref{tablesum}) show that the
values of $\lambda^{\rm ep}$ on \textit{both} the $\sigma$ sheets are approximately a factor three larger than those on
the $\pi$ sheets.

A detailed comparison with the orbit-resolved theoretical values \cite{mazin02prb} of $\lambda^{\rm ep}$ is also shown
in the table. Generally our values are slightly smaller than the theoretical ones. The most significant differences are
for the larger orbits on the $\sigma$ tubes for which $\lambda^{\rm ep}$ values which are $\sim 20\%$ smaller than
theory. The small shifts in Fermi level described above do make small differences to the calculated band masses
\cite{mazin02prb}, increasing the $\lambda^{\rm ep}_{\rm exp}$ values by $\sim$ 6\%, but the differences remain
significant. These differences could have a relatively large effect on the detailed calculations of $T_c$.  For
example, using the isotropic McMillan equation, with parameters appropriate to MgB$_2$ \cite{choi02} we estimate a 20\%
reduction in $\lambda^{\rm ep}$ would imply a 8~K reduction in $T_c$.  It is likely that the reduced $\lambda$ values
we observe are caused by phonon anharmonicity, which has been shown to reduce the average value of $\lambda$ by around
20 \%\cite{choi02}.

\begin{figure}
\includegraphics[width=7.5cm,keepaspectratio=true]{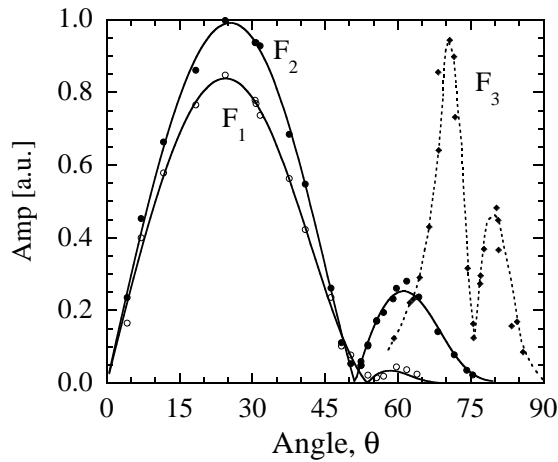}
 \caption{FFT amplitude versus angle for frequencies $F_1$, $F_2$ and $F_3$ in crystal K.  The solid lines are fits to the
 data as described in the text. The dotted line is a guide to the eye.}
 \label{figamp}
\end{figure}

Scattering rates ($\tau^{-1}$) were calculated by fitting the raw torque versus field data to Eq.\ \ref{lkexpression}.
The results are also shown in Table \ref{tablesum}. It can be seen that the mean free path ($\ell$) for sample K is
significantly longer for orbits $F_2$, $F_1$ on the $c$-axis tube but not for $F_3$ on the in-plane tubular network.
The reason for this is not clear, but it does not necessarily indicate a higher crystal quality. However, $F_5$ and
$F_6$ are almost certainly observable in crystal K because of the larger values of $\ell$ for the two $c$-axis $\sigma$
tubes.  These differences in $\ell$ may be at least partially responsible for a significant difference in $B_{c2}$
values for the two crystals. For crystal B, $B_{c2}^{\|c}$ = 3.3~T, $B_{c2}^{\bot c}$ = 17~T,  and for crystal K,
$B_{c2}^{\|c}$ = 2.5~T, $B_{c2}^{\bot c}$ = 12~T (all values quoted at $T$=0.3~K).

The longer mean free path of the smaller $\sigma$ tube ($F_1$ and $F_2$) in crystal K, has allowed us to track the
amplitude of the signal over a very wide range of angle (up to 67$^\circ$ and 81$^\circ$ respectively). The data shown
in Fig.\ \ref{figamp} show pronounced minima at $\theta=51\pm 1^\circ $, $\theta= 53\pm 1^\circ$, $\theta= 75.7\pm
0.5^\circ$ for $F_1$, $F_2$, and $F_3$ respectively, which we attribute to the spin-zero effect (the $R_s$ damping term
in Eq.\ \ref{lkexpression}). Using the calculations of $m_B(\theta)$ by Harima \cite{harima} we can deduce the
enhancement of the spin susceptibility ($1+S$) for these three orbits (at the spin zero angle). Taking $g$ = 2, we find
that $S$=0.07, 0.12 and 0.45 for the three orbits respectively.  The reason for the significantly larger enhancement on
the $\pi$ band is not clear. Band structure calculations \cite{mazin02prb} predict $S= 0.31$ and 0.26 on the $F_2$ and
$F_3$ orbits respectively.  The calculation therefore overestimates $S$ on the $\sigma$ sheet by a factor 2.5 and
underestimates it on the $\pi$ sheet by a factor 1.7.

The solid lines in Fig.\ \ref{figamp} show a fit to the data for $F_1$ and $F_2$ with Eq.\ \ref{lkexpression}, using
the $m_B$ values of of Harima \cite{harima}. There are 3 free parameters: the overall amplitude, $\tau^{-1}$ (assumed
constant as a function of angle), and $S$. The fit is remarkably good and the values of $\ell$ were found to be
410~\AA~ and 900~\AA~ for $F_1$ and $F_2$ respectively which are close to the values obtained from a fit to the field
dependence at a single angle given in Table \ref{tablesum}. However, it was  not possible to fit the data for $F_3$
because the abrupt fall for $\theta\lesssim 68^\circ$ cannot be explained simply in terms of the angular dependence of
m$_{B}$ and a constant scattering rate (Dingle plots show that $\tau$ is approximately constant with angle).

In conclusion, our data strongly support the overall topology of the predicted electronic structure of MgB$_2$ and the
calculations of the electron-phonon coupling constants for the different orbits. Our data give direct evidence that the
electron-phonon interaction is large on \textit{both} $c$-axis $\sigma$ sheets and much smaller on \textit{both}  $\pi$
sheets. We have therefore obtained conclusive evidence in favor of the two key ingredients in the two-gap model of
superconductivity in this compound, namely the Fermi surface topology and the disparity in the electron-phonon coupling
for the $\sigma$ and $\pi$ bands.

We thank a number of people for useful conversations and insights: N.W.~Ashcroft, S.~Drechsler, H.~Harima, S.M.~Hayden,
and I.~Mazin. This work was supported by the EPSRC (U.K.), NEDO (Japan) and the NSF through grant number
NSF-DMR-0084173. PJM gratefully acknowledges the support of the Royal Society (London).


\begin{thebibliography}{99}
\bibitem{harimanote} Note that $F_7$ was only predicted by one author \cite{harima},
and that this work also predicts two additional frequencies close to $F_3$ and $F_7$ which arise from slight
undulations of the Fermi surface.
\end{thebibliography}
\end{document}